\documentclass[prb,twocolumn]{revtex4-1}

\usepackage{graphicx}
\usepackage{multirow}
\usepackage{dcolumn}
\usepackage{bm}
\usepackage[normalem]{ulem}
\usepackage{amsmath}
\usepackage{amsfonts}
\usepackage{amssymb}
\usepackage{color}
\usepackage{colordvi}
\usepackage{dcolumn}

\usepackage[colorlinks=true, breaklinks=true, linkcolor=blue,
urlcolor=blue, citecolor=blue]{hyperref}

\begin{document}

\title{Optical source of individual pairs of color-conjugated photons}

\date{\today }
\author{Yury Sherkunov}
\affiliation{National Graphene Institute, University of Manchester, Manchester, M13 9PL, United Kingdom}

\author{David M. Whittaker}
\affiliation{Department of Physics and Astronomy, University of Sheffield, Sheffield, S10 2TN, United Kingdom}

\author{Vladimir I. Fal'ko}
\affiliation{School of Physics and Astronomy, University of Manchester, Manchester, M13 9PL, United Kingdom}
\affiliation{National Graphene Institute, University of Manchester, Manchester, M13 9PL, United Kingdom}
\begin{abstract}
We demonstrate that Kerr nonlinearity in optical circuits can lead to both resonant four-wave mixing and photon blockade, which can be used for high-yield generation of high-fidelity individual photon pairs with conjugated frequencies. We propose an optical circuit, which, in the optimal pulsed-drive regime, would produce photon pairs at the rate up to $10^5s^{-1}$ (0.5 pairs per pulse) with $g^{(2)}<10^{-2}$ for one of the conjugated frequencies. We show that such a scheme can be utilised to generate color-entangled photons. 

\end{abstract}

\maketitle

The use of individual photons \cite{Shields} is one of the key elements in the implementation of quantum technologies in communications security \cite{Bennett,Kimble08} and quantum computation \cite{Knill}, which stimulated a great progress in designing solid state single-photon sources \cite{Kurtsiefer00,Santori02,Wei14}. Further advancement of quantum security protocols is expected to benefit from the use of pairs of correlated photons, such as generated by the bi-exciton decay \cite{Stevenson}, spontaneous parametric down-conversion in nonlinear crystals \cite{Fasel04,Soujaeff07,Chen09,Hunault10,Fortsch}, or four-wave mixing \cite{Li05, Sharping06,Takesue07,Collins12NC,Davanco12,Sherkunov2}. To guarantee that only one photon is produced in each of the two conjugated modes, low pumping intensities had to be used in all of the above methods, leading to a low output of photon pairs \cite{Soujaeff07,Chen09,Hunault10,Fortsch,Li05,Sharping06,Takesue07,Collins12NC,Davanco12}. Here, we propose an optical circuit where the resonant four-wave mixing and photon blockade, both due to the same Kerr nonlinearity, create conditions for a high-yield generation of individual pairs of photons with conjugated frequencies. Moreover, for the pulsed pumping of such a circuit, the Rabi oscillations between different two-photon states can be used to optimise the output of such pairs, and we propose a scheme for using the optimally driven circuit as a source of color-entangled photon pairs \cite{Ramelow}.
\begin{figure}[h]
\includegraphics[width=0.4\textwidth]{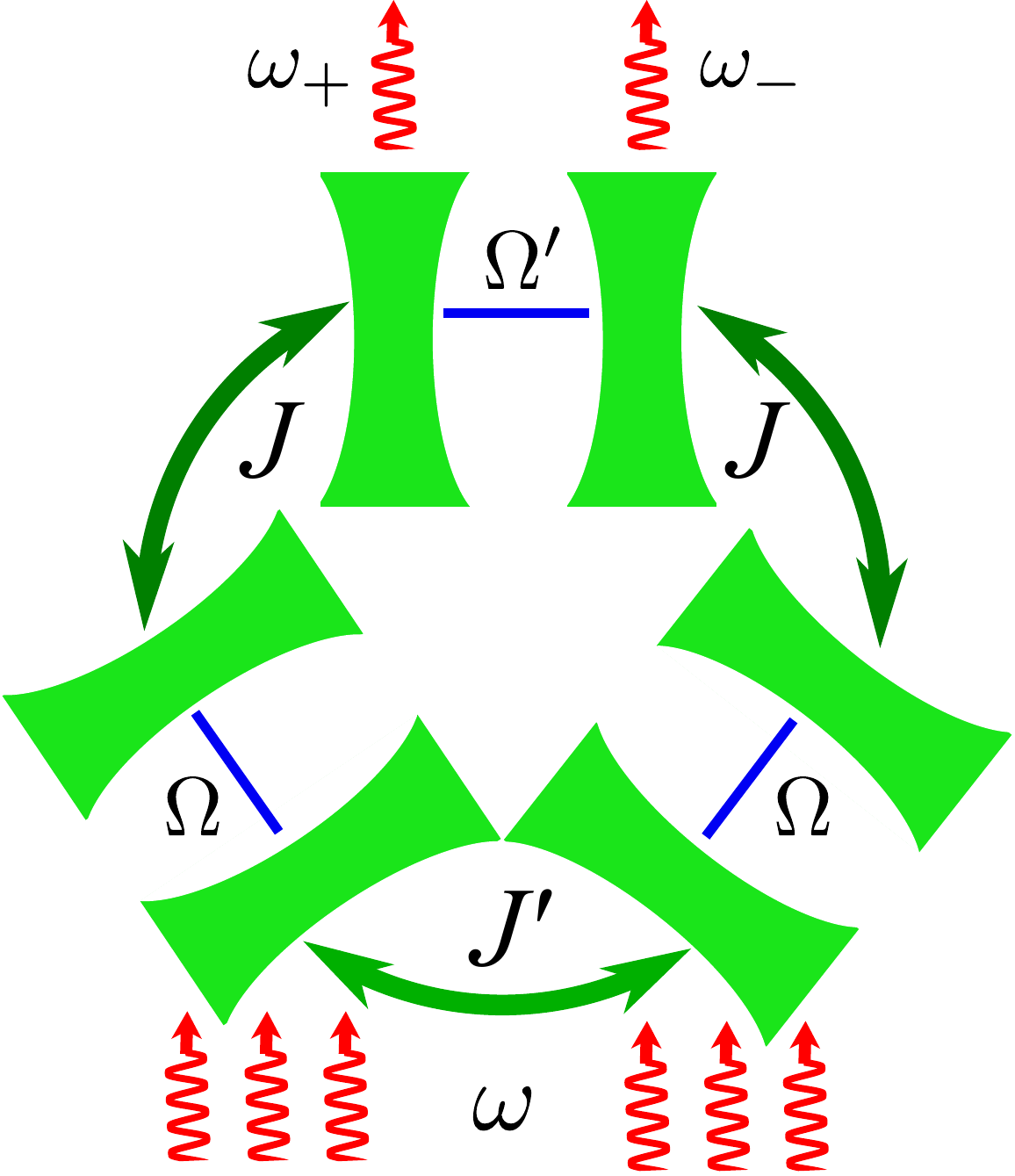}
\caption{Nonlinear optical circuit with high-Q nonlinear resonators (two with frequency $\Omega$ and one with frequency $\Omega'=\Omega+J\times \delta$, coupled by hoppings $J$ and $J'=Jx$).}  
\label{Fig1}
\end{figure} 

Spontaneous four-wave mixing (FWM) is a process that converts two photons from a coherent light source into a pair of photons with up- and down-shifted frequencies. In fibers and waveguides, such a process generates two-photon states, and additional challenge in developing devices suitable for the generation of correlated photon pairs with high fidelity and high efficiency is related to the noise due to the multi-photon generation \cite{Shields} associated with the increasing excitation power required for a high output of the device. 
 
 \begin{figure}[h]
\includegraphics[width=0.4\textwidth]{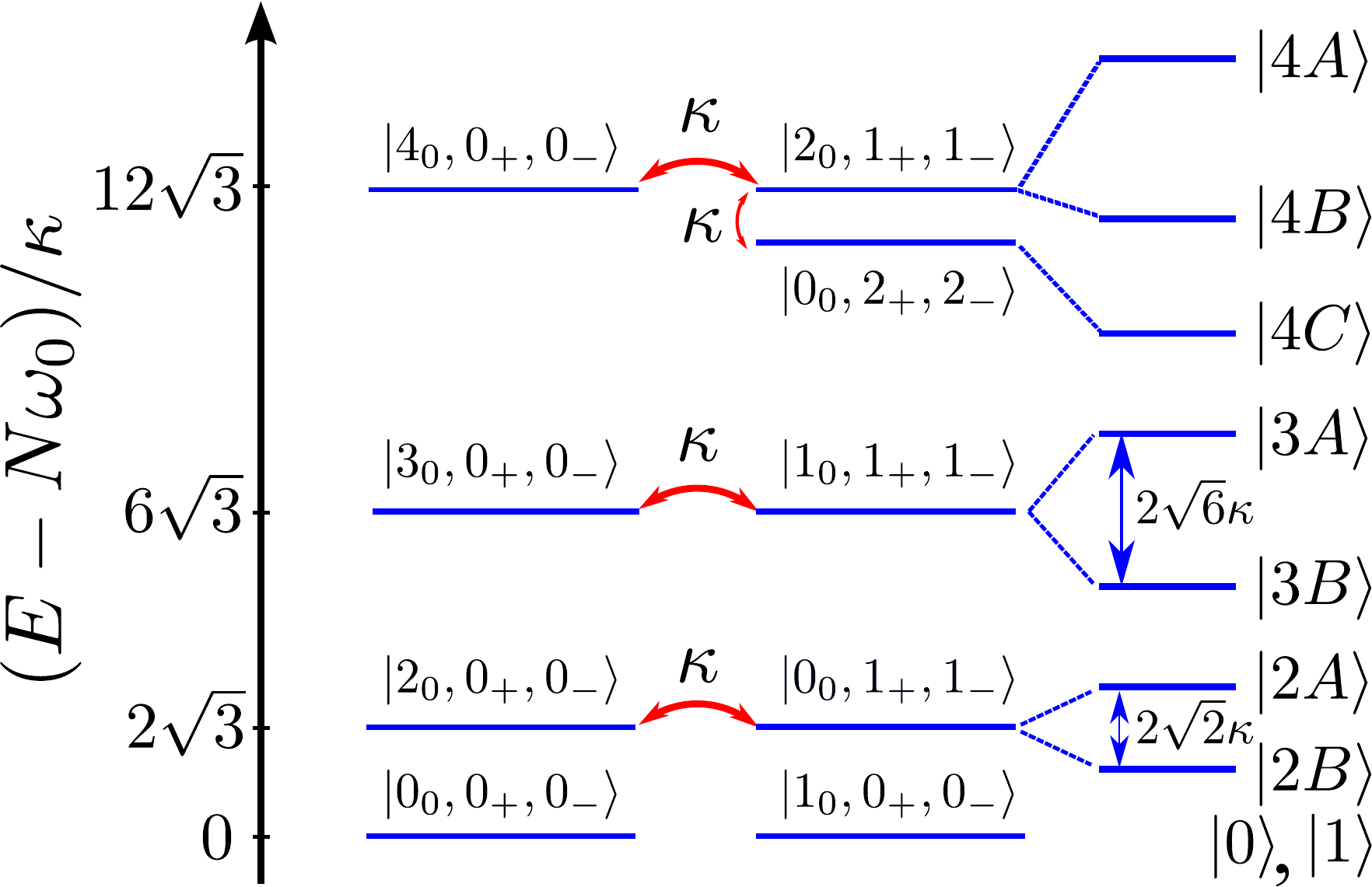}
\caption{ 
 Resonant mixing and splitting of $N$-photon states $|n_{0},n_{+},n_{-}\rangle$ coupled by the FWM described by the truncated Hamiltonian $\hat{H}^{(2)}$.  The detuning $\delta$ is chosen in such a way that the energy levels corresponding to the states $|n_{0}+2,0_{+},0_{-}\rangle$ and  $|n_{0},1_{+},1_{-}\rangle$ are  in resonance, while the energy levels of the states with multiple occupation of the modes $\omega_+$ and $\omega_-$ are red-shifted. Energies of relevant states with $N=0,1,2,3$ photons are given in Eq. (\ref{E1}).  }  
\label{Fig2}\end{figure}  

At the same time, Kerr-type nonlinearity in selectively tuned microcavity-resonators may result in emission of photon pairs in spectrally well-defined modes \cite{Sherkunov}, with the resonant conditions set by the photon blockade \cite{Verger06,Sherkunov}. Hence, we propose an optical circuit design where non-linear coupling of three photon modes with conjugated frequencies $\omega_\pm$ and $\omega_0 \approx \frac12(\omega_{+}  + \omega_{-})$ can be used for the resonant excitation of pure two-photon pairs with optimised high-yield output, followed by the Rabi-type mixing of the two-photon states $|2_{0},0_{+},0_{-}\rangle$ and $ |0_{0},1_{+},1_{-})\rangle$ (see in Fig. \ref{Fig2}), which might be used to produce entangled states of color-conjugated photon pairs.

The proposed circuit is shown in Fig. \ref{Fig1}. It can be envisaged as three coupled non-linear cavity resonators, each characterised by a single photonic mode: two with equal frequencies, $\Omega$, and  the third with the frequency $\Omega'=\Omega+J\times\delta$.   The two cavities with equal frequencies are coupled with the third cavity by hopping amplitude $J$ and by hopping amplitude $J'=Jx$ with each other, as described by the Hamiltonian ($\hbar=c=1$):
\begin{eqnarray}
\hat{H}^{(0)}&=&\Omega' a_1^{\dagger}a_1+\Omega \sum_{i=2,3} a_i^{\dagger}a_i \label{BHtr}\\
&+&J' a_2^{\dagger}a_3+J\sum_{i=2,3}a_1^{\dagger}a_i+H.c.
\equiv\sum_{k=0,\pm}\omega_{k}\beta_k^{\dagger}\beta_k,\nonumber\\
\omega_{0}&=&\Omega-Jx,\;\; \omega_{\pm}=\Omega+\frac{J}{2}(\delta+x\pm s),\nonumber\\
s&=&\sqrt{(\delta-x)^2 +8}\nonumber,
\end{eqnarray}
where $a_i$ ($a_i^{\dagger}$) are the annihilation (creation) operators of  photons in each resonator.  Energies  $\omega_{0,\pm}$ correspond to extended eigenmodes of  $H^{(0)}$, 
\begin{eqnarray}
\beta_0&=&\frac{1}{\sqrt{2}}\left(a_3-a_2\right),\;\;c_{\pm}=\delta-x\pm s,\nonumber\\
\beta_{\pm}&=& \frac{1}{\sqrt{8+c_{\pm}^2}}\left[c_{\pm}a_1+2(a_3+a_2)\right].
\end{eqnarray}
Resonators 2 and 3 are driven by a coherent pump (at frequency $\omega$) with the amplitudes $\pm \sqrt 2F$, which provides coupling to the mode $\beta_0$, 
\begin{eqnarray}
\hat{H}^{(1)}=-Fe^{i\omega t}\beta_0+H.c.\label{HptrN}
\end{eqnarray}

Then, we take into account Kerr-type nonlinearity, which, for simplicity, will have the same strength, $u\ll J$, on each cavity,  
\begin{eqnarray}
\hat{H}^{(2)} &=& u\sum_{i}a_{i}^{\dagger}a_i^{\dagger}a_ia_{i}\label{Hnltr}\\
&=& \kappa\left(\beta_+^{\dagger}\beta_-^{\dagger}\beta_0^2+H.c. \right)+ \sum_{k,k'}^2\alpha_{kk'}\beta_k^{\dagger}\beta_{k'}^{\dagger}\beta_{k'}\beta_k + \delta \hat{H}, \nonumber
\end{eqnarray}
where
\begin{eqnarray}
\kappa&=&\frac{\sqrt{2}u}{s},\;\;\alpha_{00}=\frac{\kappa s}{2\sqrt 2}, \;\;
\alpha_{\pm\mp}=\frac{6\kappa}{\sqrt{2} s},\nonumber\\
\alpha_{\pm0}&=&\alpha_{0\pm}=\frac{\kappa}{2 \sqrt{2}}[s\mp(\delta-x)], \nonumber\\
\alpha_{\pm \pm}&=&\frac{\kappa}{4 \sqrt 2 s}[3s^2-12\pm(\delta-x)s]. \nonumber
\end{eqnarray}
The second line in $\hat{H}^{(2)}$ represents interaction between the extended modes, $\beta_{0,\pm}$. Here, the first term describes resonant four-wave mixing of two $\omega_0$ photons with the pair of photons at the conjugated frequencies, $\omega_\pm$. The second term produces occupancy-dependent shifts in the photon frequencies. The rest of the terms generated by the canonical transformation from the single-cavity to the extended modes are combined into a perturbation $\delta \hat{H}$; under conditions which will be identified below, these terms are non-resonant for the production process of the photon pairs with conjugated frequencies, hence, they give only a small contribution (but they will be included in numerical calculations later).

The 'resonant' situation occurs when the two states 
\begin{eqnarray}
&|n_{0}+2,n_{+},n_{-}\rangle&=\frac{(\beta_0^{\dagger})^{n_0+2}(\beta_+^{\dagger})^{n_+}(\beta_-^{\dagger})^{n_-}}{\sqrt{(n_0+2)!n_+!n_-!}}|0\rangle,\nonumber\\
&|n_{0},n_{+}+1,n_{-}+1\rangle&=\frac{(\beta_0^{\dagger})^{n_0}(\beta_+^{\dagger})^{n_++1}(\beta_-^{\dagger})^{n_-+1}}{\sqrt{n_0!(n_++1)!(n_-+1)!}}|0\rangle\nonumber
\end{eqnarray}
are degenerate,
\begin{eqnarray}
E(n_{0}+2,n_{+},n_{-})=E(n_{0},n_{+}+1,n_{-}+1).\label{conserv}
\end{eqnarray}
In this case, generation of pairs of $\omega_\pm$-photons is promoted by the resonance conditions for converting them from the pairs of pumped $\omega_0$ photons. Condition (\ref{conserv}) can be obtained by tuning parameter $\delta$ in $H^{(0)}$ to the value 
\begin{eqnarray}
\delta &=&-3x+\frac{ \kappa}{2J}\left[\frac{4x^2-1}{\sqrt{1+2x^2}}(2+n_++n_-)\right.\nonumber\\
&\;\;&\;\;\;\;\;\;\;\;\;\;\;\;\;\;\;\;\;\;\;\;\;\;\;\;\;\;+\left.10\sqrt 2 x(n_+-n_-)\right]. \nonumber
\end{eqnarray}
The latter expression was obtained neglecting terms $\delta \hat{H}$ in $\hat{H}^{(2)}$. This indicates that the resonance conditions for the states involving different occupation numbers $n_{\pm}$ can be separated, whereas the resonance conditions for the processes $|n_0+2,0_+,0_-\rangle\Longleftrightarrow |n_0,1_+,1_-\rangle$ generating a single pair of photons at conjugated frequencies $\omega_{\pm}$,
\begin{equation}
\delta=-3x+\frac{4x^2-1}{4x^2+2}\frac{u}{J} ,\label{dw}
\end{equation}
is the same for all values of $n_0$. This condition sets the values of the parameters in Eqs. (\ref{BHtr}) and (\ref{Hnltr}) to 
\begin{eqnarray}
\omega_{0}&=&\Omega-Jx;\;  \kappa \approx \frac{u}{2\sqrt{1+2x^2}};\; \label{hparamres}\\
\omega_{\pm}&=&\Omega-\left(x\mp \sqrt 2 \sqrt{1+2x^2}\right)J\nonumber\\
&\;\;&+\frac{4x^2-1}{4x^2+2}\left( \sqrt{1+2x^2}\mp\sqrt 2 x\right)\kappa;\nonumber\\
\alpha_{00}&=& \sqrt{1+2x^2}\kappa; \;  \alpha_{\pm\pm}=\frac{3+12x^2\mp  2x\sqrt{2+4x^2}}{4 \sqrt{1+2x^2}}\kappa;\nonumber\\
\alpha_{\pm 0}&=&\alpha_{0\pm}=[\sqrt{1+2x^2}\pm\sqrt 2 x]\kappa; \; \alpha_{\pm\mp}=\frac{ 3\kappa}{2\sqrt{1+2x^2}}.\nonumber
\end{eqnarray}

To get an idea about spectral properties of this optical circuit under the resonance conditions (\ref{dw})-(\ref{hparamres}), we neglect the term $\delta \hat H$ in $H^{(2)}$ and diagonalise   
\begin{equation}
\hat H=\hat H^{(0)}+\hat H^{(2)} \label{H}
\end{equation}
in the basis  of the states $\{|0_{0}, 0_{+},0_{-}\rangle$, $|1_{0}, 0_{+},0_{-}\rangle$, $|2_{0}, 0_{+},0_{-}\rangle$, $|3_{0}, 0_{+},0_{-}\rangle$, $|0_{0}, 1_{+},1_{-}\rangle$, $|1_{0}, 1_{+},1_{-}\rangle\}$. Formally, truncation decouples this subspace of states from the rest of the low-energy spectrum with the same total number of photons. Then, the spectrum of $N-$photon states is: 
\begin{eqnarray}
E(0)&=&0,\;\;E(1)=\omega_0, \nonumber\\
|0\rangle&=&|0_{0},0_{+},0_{-}\rangle,\;\;|1\rangle=|1_{0},0_{+},0_{-}\rangle ;\nonumber\\
E(2\alpha)&\approx&2\omega_0+\left(\sqrt{2+4x^2}\pm 1\right)\sqrt 2\kappa ,\label{E1}\\
|2\alpha\rangle &=&\frac{1}{\sqrt 2}(|2_{0},0_{+},0_{-}\rangle \pm |0_{0},1_{+},1_{-})\rangle; \nonumber\\
E(3\alpha)&\approx&3\omega_0+\left(\sqrt{6+12x^2}\pm 1\right)\sqrt 6\kappa,\nonumber\\
|3\alpha\rangle &=&\frac{1}{\sqrt 2}(|3_{0},0_{+},0_{-}\rangle\pm |1_{0},1_{+},1_{-}\rangle). \nonumber
\end{eqnarray}  
Here, upper/lower signs correspond to the states $A/B$ among $|N\alpha\rangle_{N=0,1,2,3;\alpha=A,B}$ marked in Fig. \ref{Fig2}.

\begin{figure}[h]
\includegraphics[width=0.5\textwidth]{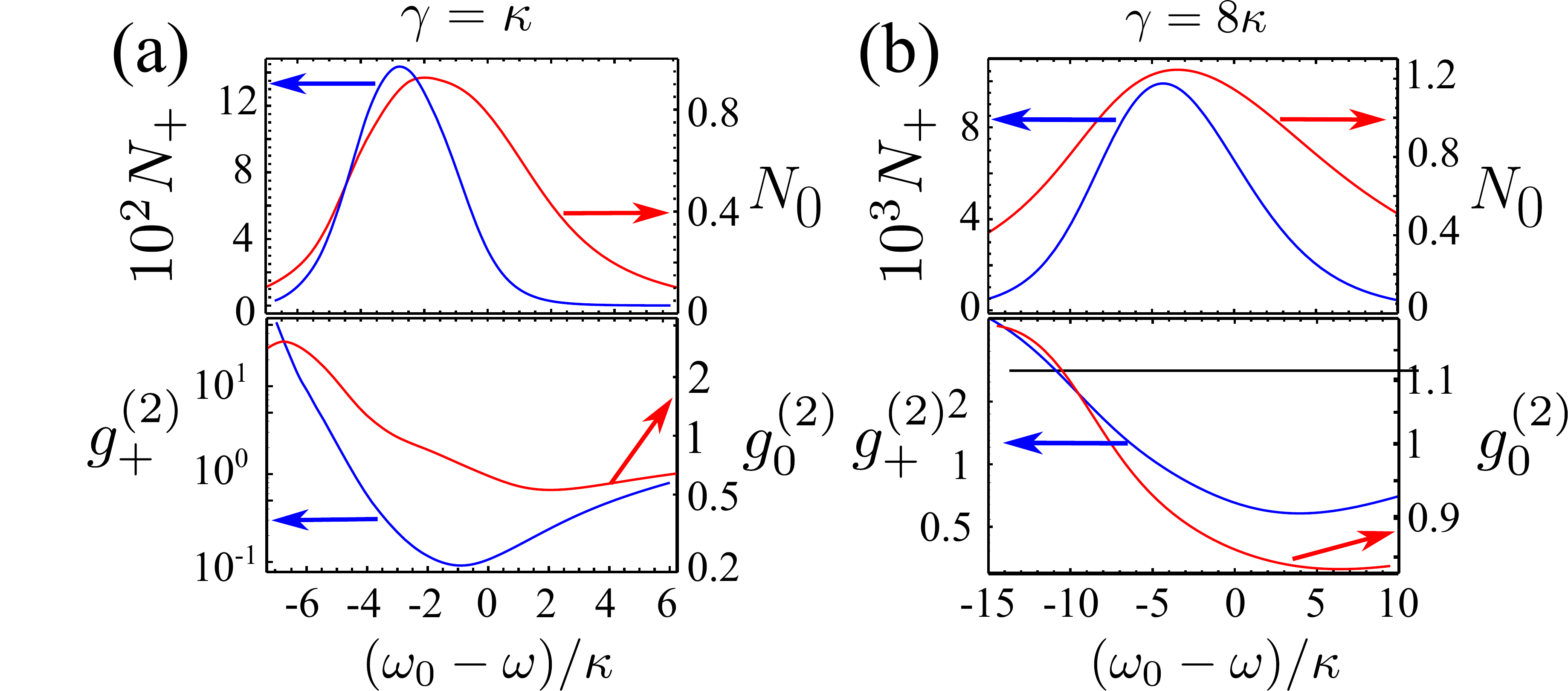}
\caption{ Steady state under continuous pumping, as a function of pump frequency $\omega$.    (a) $\gamma=\kappa$, $F=2\kappa$; (b) $\gamma=8\kappa$, $F=10\kappa$.     }   
\label{Fig3}
\end{figure}

In a pumped system, absorption of photons would be resonantly favoured when $N$ incident photons have the same energy as $N$ photons in the cavity,
$N\omega=E(N\alpha)$, and such multi-photon resonances can also be found in weakly dissipating systems. 
The evolution of a quantum optical circuit, where photons experience decay at the rate $\gamma$, can be described using the master equation, 
\begin{equation}
\frac{\partial\hat{\rho}}{\partial t}=-i[\hat H,\hat{\rho}]+\gamma \sum_{i=0,1,2}(2 \beta_i\hat{\rho}  \beta_i^{\dagger}-\beta_i^{\dagger}  \beta_i\hat{\rho} -\hat{\rho}  \beta_i^{\dagger} \beta_i), \label{mastereq}
\end{equation} 
for the density matrix, $\rho$, which we write in a Fock basis
\begin{eqnarray}
\hat{\rho}=\sum \rho(m_{0},m_{+},m_{-};n_{0},n_{+},n_{-}) \nonumber \\
 |m_{0},m_{+},m_{-}\rangle \langle n_{0},n_{+},n_{-}|. \nonumber
\end{eqnarray}  
By solving  this master equation numerically using the basis that includes states with up to $N_{max}\le 10$ photons in each mode,
we calculate the occupation numbers $N_i=\mbox{Tr} [\beta_i^{\dagger}\beta_i\hat{\rho}]$, zero time-delay pair correlation functions for each mode $\omega_{0,\pm}$, $g^{(2)}_i=\mbox{Tr} [(\beta_i^{\dagger})^2\beta_i^2\rho]/N_i^2$
and the probabilities $P(2_0)$ and $P(1_{+},1_{-})$  to find a photon pair in the mode $\omega_{0}$ or  the conjugated modes, respectively. We checked the consistency of our calculations by converging the results upon increasing $N_{max}$ and by comparing the results of modelling where we include and neglect the interaction terms $\delta \hat{H}$.   

\begin{figure}[h]
\includegraphics[width=0.3\textwidth]{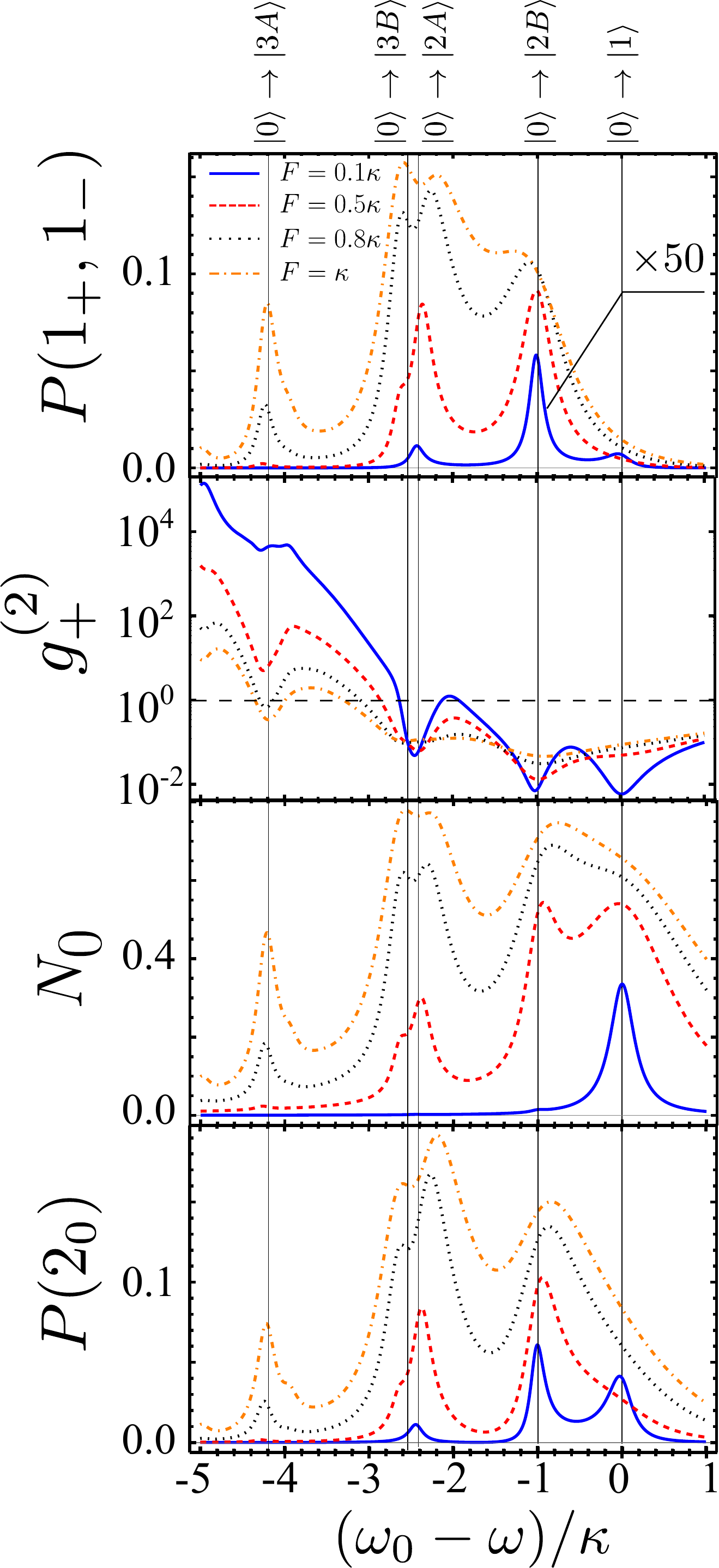}
\caption{ Dependence of steady state values of $P(1_+,1_-)$, $g^{(2)}_+$, $N_0$ and $P(2_0)$ on the pump frequency $\omega$ for $\gamma=0.1\kappa$ and various amplitudes of the pump. Relevant resonance conditions correspond to multi-photon transitions sketched in Fig. \ref{Fig2}.   }  
\label{Fig4}
\end{figure}

\begin{figure}[h]
\includegraphics[width=0.4\textwidth]{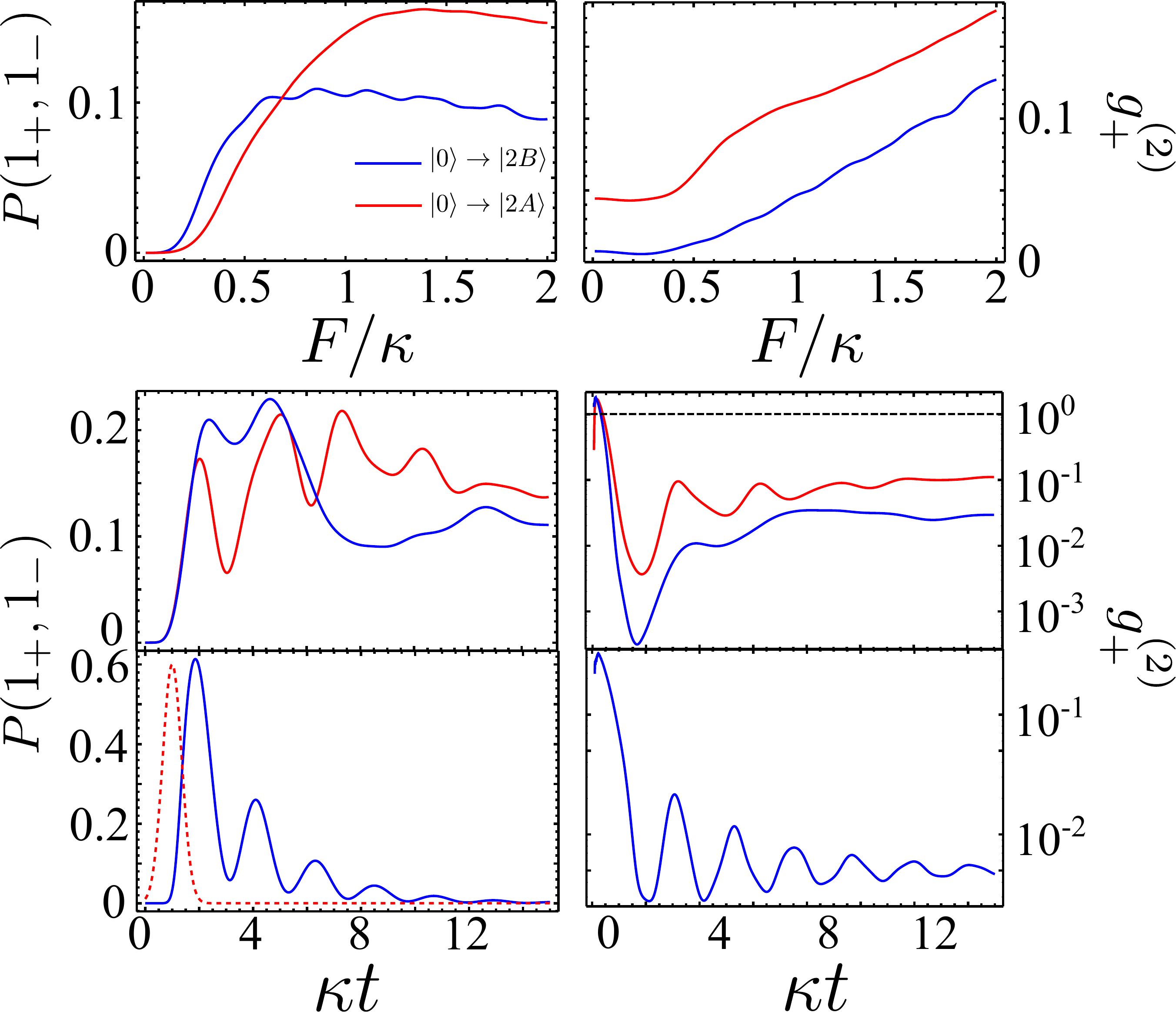}
\caption{  Top: Steady-state parameters computed in the system with $\gamma=0.1\kappa$ under continuous pumping as a function of pumping amplitude $F$, for resonance frequencies of $|0\rangle\rightarrow|2A\rangle$ and $|0\rangle\rightarrow|2B\rangle$ transitions marked on Fig. \ref{Fig4}. Middle: Time evolution of pumped circuit, following the switching-on of the pump $F=0.8\kappa$. Bottom: Rabi-type oscillations of two-photon states in a circuit pumped by a Gaussian pulse (dashed line) with parameters indicated on Fig. \ref{Fig6}.}   
\label{Fig5} 
\end{figure}

First, we solved Eq. (\ref{mastereq}) for a circuit with $J'=J$ ($x=1$) and $\gamma=\kappa$ or $\gamma=8\kappa$ (which is typical for GaAs polaritonic microcavities \cite{Dufferwiel14}), with weak anti-bunching in the low-flux of $\omega_\pm$ photons demonstrated by $g^{(2)}_+$ shown in Fig. \ref{Fig3}. In contrast, for the resonant conditions set for $J'=J$ ($x=1$) and with $\gamma=0.1 \kappa$ in a system  continuously pumped with amplitude $F$, we find a much higher efficiency of production of pure two-photon states. The numerically found steady-state solutions of Eq. (\ref{mastereq}) displays resonances corresponding to the transitions in the spectrum in Fig. \ref{Fig2}. This is illustrated in Fig. \ref{Fig4} by the pump-frequency $\omega$ dependence of probabilities $P(1_+,1_-)$ and $P(2_0)$ to find one $\omega_\pm$ pair or two $\omega_0$ photons in the circuit, occupation numbers for the excited $\omega_0$ photons, and the two-photon correlation function $g_+^{(2)}$. For $F\ll\kappa$, $P(2_0)$ and $P(1_+,1_-)$ have pronounced resonances at  $2\omega=E(2A)$ and $2\omega=E(2B)$, corresponding to the two-photon transitions $|0\rangle\rightarrow|2A\rangle$ and $|0\rangle\rightarrow|2B\rangle$. Resonant excitation of individual $\omega_\pm$ pairs is also reflected by the dips in $g_+^{(2)}$. For larger $F$, we identify additional resonances in the vicinity of $3\omega=E(3A)$ and $3\omega=E(3B)$, corresponding to the three-photon transitions $|0\rangle\rightarrow|3A\rangle$ and $|0\rangle\rightarrow|3B\rangle$. Note that, at a larger $F$, the pump induces shifts in the resonance conditions to create multi-photon states, Eq. (\ref{E1}), and the maxima, \textit{e.g.} in $N_0$ shown in Fig. \ref{Fig4}, are additionally shifted by power-dependent broadening of $|0\rangle\rightarrow|1\rangle$ resonance. 

As the pumping amplitude grows, the two-photon resonances $|0\rangle\rightarrow|2A\rangle$ and $|0\rangle\rightarrow|2B\rangle$ become more pronounced, accompanied by increasing $P(1_{+},1_{-})$. However, with the reference to Fig. \ref{Fig5}, where two top panels show the steady state parameters for the circuit pumped at the resonance frequencies of $|0\rangle\rightarrow|2A\rangle$ and $|0\rangle\rightarrow|2B\rangle$ transitions, saturation of $P(1_{+},1_{-})$ at $F\sim\kappa$, accompanied by pollution of photon pairs at conjugated frequencies with individual $\omega_\pm$ photons (manifested by increasing $g^{(2)}_+$), suggest that a mere increase of pumping does not improve the output of correlated photon pairs.

\begin{figure}[h]
\includegraphics[width=0.4\textwidth]{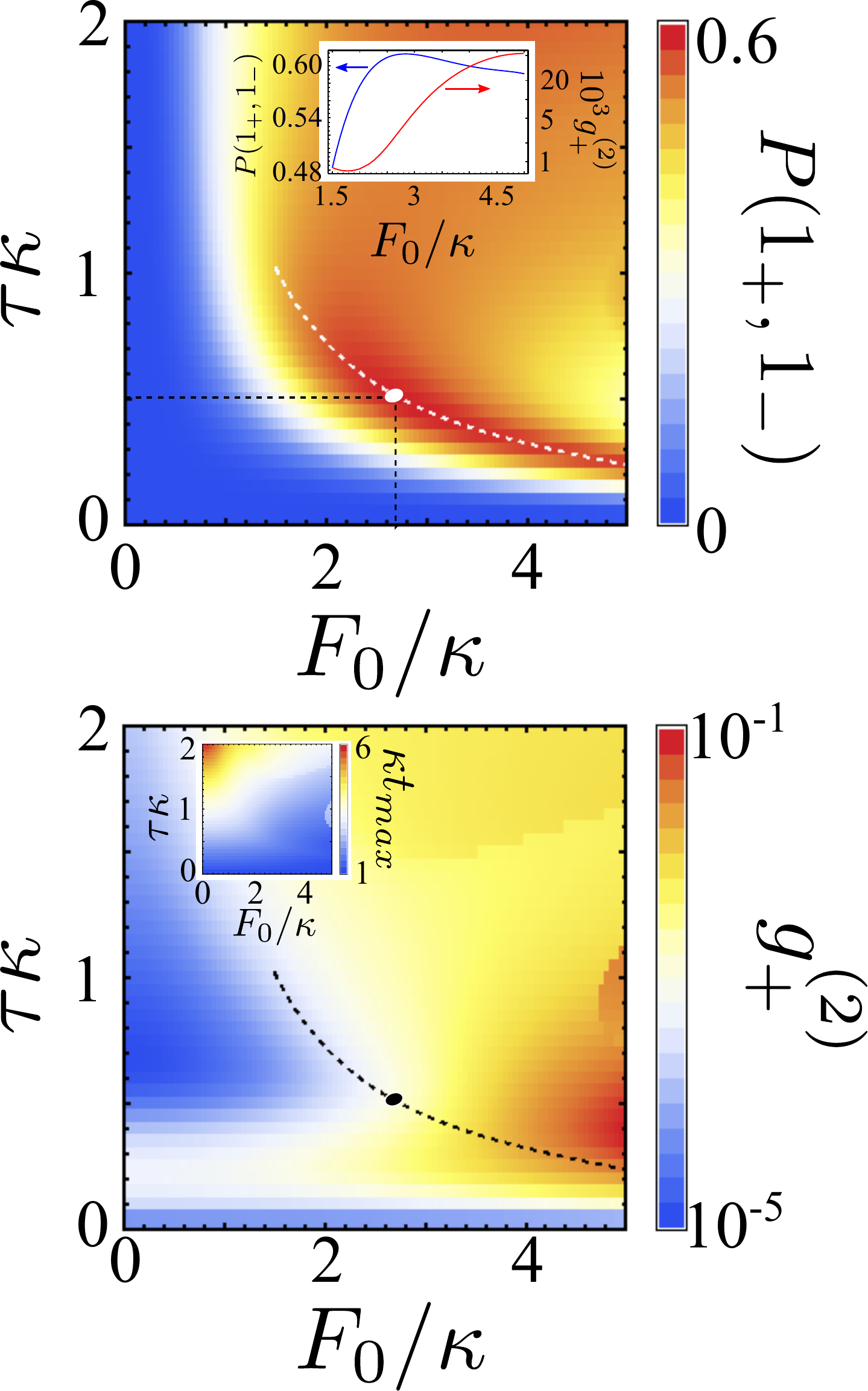}
\caption{ $P(1_{+},1_{-})$ (top) and $g_+^{(2)}$ (bottom) at $t=t_{max}$ for the system pumped by a Gaussian pulse with amplitude $F=F_0e^{-(t-2\tau)^2/\tau^2}\theta(t)$. White (black) dot marks a favourable choice of $\tau$ and $F_0$ used in Fig. \ref{Fig5}. Insets: (top) correlated photon pair output and $g_+^{(2)}$ along the line of maximal $P(1_{+},1_{-})$   at $t=t_{max}$ and (bottom) parametric dependence of $t_{\max}\kappa$. }   
\label{Fig6} 
\end{figure}

The insight into how one can increase the output of individual photon pairs comes from the pronounced beatings in the temporal evolution of $P(1_{+},1_{-})$ and $g_+^{(2)}$, which follow switching-on of the excitation source  $F=\theta(t)\times \mathrm{const}$ ($\theta$ is the Heaviside function), Fig. \ref{Fig5}. These beatings are the results of two-photon Rabi oscillations \cite{Sherkunov1}, generated by resonance mixing of $|n_0+2,0_+,0_-\rangle\Longleftrightarrow |n_0,1_+,1_-\rangle$ states. Hence, we suggest to implement pulsed excitations, harvesting photon pairs within optimally chosen delay-time windows. Note that, for $\kappa \ll J$, the period, $\pi/\sqrt{2}\kappa$, of Rabi oscillations $|2,0_+,0_-\rangle\Longleftrightarrow |0,1_+,1_-\rangle$, is long enough for harvesting correlated photon pairs at the time interval around the optimal delay $t_{max}$ at the maximum of $P(1_{+},1_{-})$, without undermining their spectral identity. Hence, we identify time intervals of the maximal probability to find a high-fidelity conjugated photon pair (in those intervals, $N_+ \approx P(1_{+},1_{-})$). An example of time-dependent $P(1_{+},1_{-})$ and $g_+^{(2)}$, produced by a Gaussian pulse of duration $\tau$, is shown in the bottom panels in Fig. \ref{Fig5}, and in Fig. \ref{Fig6} we show the dependence of the size of the maximum output $P(1_{+},1_{-})$, at $t_{max}$, and $g_+^{(2)}\sim10^{-3}$ . The optimal choice of the duration and amplitude of the pulse offers a high yield, $P(1_+,1_-) \sim 0.5$ of an almost pure two-photon state.      
 
To achieve the desirable regime of $\kappa/\gamma\gg 1$, one needs to use materials with a large non-linearity and cavities with a high quality factor, $Q$. Depending on the operational frequency range, these may be $Q\sim 10^9$ superconducting microwave resonators \cite{Barends10,Geerlings12}, coupled with superconducting qbits for microwave frequencies, or trapped ions in the electromagnetically-induces-transparency regime \cite{Imamo97, Hartmann07} resonantly coupled to $Q\sim 10^{7-8}$ toroidal \cite{Armani03} or microrod \cite{Del13} microcavities for visible or infrared frequencies. In the latter systems\cite{Armani03,Aoki06,Hartmann07}, non-linearity can reach $\kappa\sim 1.25 \times 10^7 s^{-1}$, with $\gamma\sim 2\times 10^5 s^{-1}$, and the optimised pulses and repetition rate 
$0.1\gamma$ would produce pairs of colour-conjugated photon pairs with $g_s^{(2)} \sim 10^{-2} $ at the rate of up to $10^5 s^{-1}$, much higher than what has been achieved using the parametric down-conversion \cite{Fortsch} (pairs with $g_s^{(2)}\sim 10^{-1}$ at the rate of rate $10^2 s^{-1}$).   

Finally, the proposed non-linear optical circuit can be used as a color-entangled photon source \cite{Ramelow} by connecting one waveguide L to resonator 1 and another waveguide R equally coupled to both resonators 2 and 3 used for the excitation pulse. The escape of the two-photon state $|1_{+},1_{-}\rangle$ into the waveguides L/R, with couplings $\sim\gamma$, would deliver signals to the recipients at the L and R ends, 
\begin{eqnarray}
|1_{+},1_{-}\rangle &\rightarrow& \left[\frac{1}{2} - \frac{x}{\sqrt 2 \sqrt{1+2x^2}} \right]   |L_{+},R_{-}\rangle \nonumber \\ 
&-& \left[\frac{1}{2} + \frac{x}{\sqrt 2 \sqrt{1+2x^2}} \right]  |R_{+},L_{-}\rangle  \label{Bell} \\
&+& \frac{1}{2\sqrt{1+2x^2}} \left[ |R_{+},R_{-}\rangle - |L_{+},L_{-}\rangle\right],\nonumber
\end{eqnarray}
who would detect arrival of photons, distinguishing their color. Projecting the wave function  Eq.(\ref{Bell}) onto the subspace with one photon at L and one at R, recipients L and R would be able to use the color-entangled photon pairs similarly to what was suggested for the polarisation-entangled photon pairs \cite{Shields,Stevenson,Chen09}. Then, optimal output of the color-entangled states would be achieved in a circuit with $x \rightarrow 0$ ($J' \ll J$, corresponding to a simple linear chain of three cavity-resonators with $\omega_\pm \approx \omega_0 \pm \sqrt{2} J$, well separated from both  $\omega_0$ mode and the pumping field), for which the two-photon states in the first two lines of Eq. (\ref{Bell}) would be nothing but a Bell pair.

{\bf Acknowledgements}
We thank P. Kok, S. Flach,  M. Skolnick, and L. Glazman for useful discussions. This
work was supported by EPSRC Programme Grant
EP/J007544.

\end{document}